# Uniform focusing of sequence of relativistic positron bunches in plasma

*Vasyl Maslov (1, 2), Denys Bondar (1, 2), Iryna Levchuk (1), Sofiia Nikonova (2),*
*Ivan Onishcenko (1)*
*((1) National Science Center "Kharkov Institute of Physics and Technology", Kharkov, Ukraine*
*(2) V.N. Karazin Kharkov National University, Kharkov, Ukraine)*

vmaslov@kipt.kharkov.ua    bondar@kipt.kharkov.ua

Plasma-based accelerators sustain accelerating gradients which are several orders greater than obtained in conventional accelerators. Focusing of electron and positron beams by wakefield, excited in plasma, in electron-positron collider is very important. The focusing mechanism in the plasma, in which all electron bunches of a sequence are focused identically, has been proposed by authors earlier. The mechanism of focusing of a sequence of relativistic positron bunches in plasma, in which all positron bunches of sequence are focused identically and uniformly, has been investigated in this paper by numerical simulation by 2.5D code LCODE. Mechanism of this identical and uniform focusing involves the use of wave-length $\lambda$, which coinciding with double longitudinal dimension of bunches $\lambda=2\Delta_b$, the first bunch current is in two times smaller than the current of the following bunches of sequence and the distance between bunches equals to one and a half of wavelength 1.5$\lambda$. We numerically simulate the self-consistent radial dynamics of lengthy positron bunches in homogeneous plasma. In simulation we use the hydrodynamic description of plasma. In other words, the plasma is considered to be cold electron liquid, and positron bunches are aggregate of macroparticles. Positron bunches are considered to be homogeneous cylinders in the longitudinal direction. Positrons in bunches are distributed in radial direction according to Gaussian distribution. It is shown that in this case only first bunch is in the finite longitudinal electrical wakefield $E_z\neq 0$. Other bunches are in zero longitudinal electrical wakefield $E_z=0$. Between bunches of this sequence longitudinal electrical wakefield and radial force are not zero $E_z\neq 0$, $F_r\neq 0$. The focusing radial force in regions, occupied by bunches, is constant along each bunch $F_r$=const. Between bunches the radial force is inhomogeneous $F_r\neq$const. All positron bunches of sequence are focused identically and uniformly.

Plasma-based accelerators sustain accelerating gradients which are several orders greater than obtained in conventional accelerators [1-3]. Accelerating wakefield can be excited by single electron bunch [4, 5]. As plasma is inhomogeneous and nonstationary it is difficult to excite wakefield resonantly by a long sequence of electron bunches [6, 7], to focus sequence [8-12], to prepare sequence from long beam [13-15] and to provide large transformer ratio [16-22]. In [7] the mechanism has been found and in [23-26] investigated of resonant plasma wakefield excitation by a nonresonant sequence of short electron bunches.

Focusing of electron and positron beams by wakefield, excited in plasma, in electron-positron collider is very important [8-10, 27-30]. The focusing mechanism in the plasma, in which all electron bunches of a sequence are focused identically, has been proposed in [8-10]. However, investigations show that in a strongly nonlinear regime the value and spatial distribution of wakefield, excited by sequence of positron bunches, are different in comparison with the value and spatial distribution of wakefield, excited by sequence of electron bunches. Therefore this lens for relativistic positron bunches is researched in this paper by numerical simulation by 2.5D code LCODE [31]. Code LCODE treats plasma as a cold electron fluid and the bunches as ensembles of macro-particles. Electron beam is represented by a sequence of 4 and 10 electron bunches.

The article deals only with the focusing process. This regime can take place in the following cases:
-before the meeting point of colliding beams;
- during beam transport;
-in conditions with spatially separated processes of acceleration and focusing.

We use the cylindrical coordinate system ($r$, $z$) and plot wakefields, plasma and beam densities at some $z$ as functions of the dimensionless value $\xi=(z-V_b t)$, $V_b$ is the velocity of bunches.

Wakefield is normalized on $E_0=cm\omega_p/e$, where m is the electron mass, e is the elementary charge, c is the speed of light, and $\omega_p=(4\pi n_0 e^2/m)^{1/2}$ is the plasma frequency. Time t is normalized on $\omega_{pe}^{-1}$, longitudinal momentum of bunches $P_z$ – on $mc\gamma_b$, radius of bunches on $c/\omega_p$, beam current $I_b$ – on $mc^3/e$, emittance of bunches $\sigma$ - on $c^2/\omega_p$; plasma electron density $n_e$ and bunch density $n_b$ are normalized on unperturbed plasma electron density $n_0$, radial r and longitudinal z coordinates - on $c/\omega_p$.

All bunches of sequence are focused identically and uniformly. By code LCODE we simulate the behavior of positron bunches of finite dimension in uniform plasma. The code simulate plasma electrons, using the hydrodynamic equations. In other words the plasma electrons are considered to be cold electron liquid, and bunches are aggregate of macroparticles.

Aim of paper is to show that all relativistic positron bunches of sequence can be focused identically and uniformly and to derive conditions for achievement of identical and uniform focusing of relativistic positron bunches of sequence.

## RESULTS OF SIMULATION

Let us study focusing of positron bunches of sequence by wave, which wave-length coinciding with double longitudinal dimension of bunches. This case is interesting, since at growth of bunch longitudinal dimension at fixed its current the amplitude of wakefield reaches highest value at $\lambda=2\Delta_b$. We apply the first bunch current is in two times smaller than the current of the following bunches of sequence $I_1=I/2$, $I_i=I$, $i=2, 3, 4 ...$, spatial dimension from one bunch to another coincides with $1.5\lambda$, $\lambda$ is used wave-length of wakefield. In this case the distribution of excited longitudinal wakefield $E_z$, radial wake force $F_r$ and magnetic field $H_\theta$ are of the form, shown in Fig. 1 for four bunches.

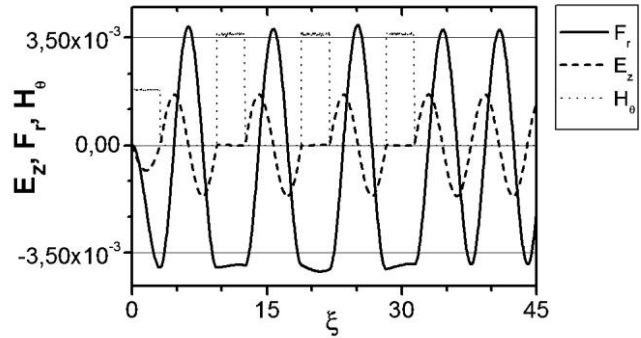

Fig. 1. Off-axis longitudinal wakefield $E_z$ (- - -), off-axis radial wake force $F_r$ (----)=$E_r$-$V_b H_\theta/c$ and off-axis magnetic field $H_\theta$ (······) for $z=3$, $\gamma_b=5$, $I_b=0.3\times10^{-3}$, $r_b=0.1$, $\gamma_b$ is the relativistic factor of bunches, $I_b$ is the maximal beam current. $E_z$, $F_r$, $H_\theta$ have been calculated for radius $r=r_b$

In Fig. 1 one can see positions of bunches by positions $H_\theta$, because magnetic field is created by beam current. From Fig. 1 it is evidently, that first bunch is in $E_z\neq 0$ and it excites wakefield. All following bunches are in $E_z=0$ and they do not excite wakefield. Hence wakefield does not change from one bunch to another. However the sequence is focused since amplitude of transversal wake force is finite $F_r\neq 0$ in areas of bunches.

It is necessary to list the distinctive characteristics of the lens of this mechanism:

1) radial wake force $F_r$ does not approximately depend on coordinate in regions, occupied by bunches (with the exception of first bunch), $F_r\approx$const, i.e. lengthy bunches are focused identically;
2) only first bunch is decelerated;
3) identical focusing force effects on all bunches (with the exception of first bunch);
4) longitudinal wakefield equal zero $E_z=0$ in regions, occupied by bunches (with the exception of first bunch).

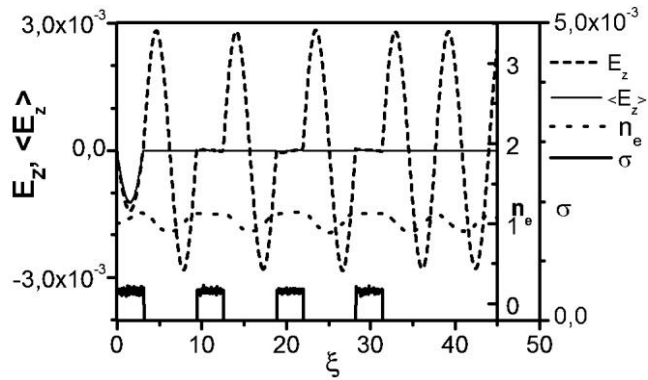

Fig. 2. On-axis plasma electron density $n_e$ (······) in wakefield, emittance of bunches $\sigma$ (------), on-axis longitudinal wakefield $E_z$ (- - -) and $<E_z>=\int dr\, r\, E_z n_b/\int dr\, r n_b$ (------) coupling rate of bunch with wakefield $E_z$ for $z=3$, $\gamma_b=5$, $I_b=0.3\times10^{-3}$, $r_b=0.1$

Such ideal focusing (Fig. 3) is realized due to formation of flat elevations of plasma electron density $n_e$ in areas of bunches (Fig. 2). These flat elevations compensate charges of bunches but magnetic field of current of bunches focuses them. In first bunch area (Fig. 2), the excited elevation of plasma electron density $n_e$ is not uniform. As a result first bunch nonuniform focusing is developed (Fig. 3). It is evidently from Fig. 2 that perturbation of density of plasma electrons $n_e$ is periodic but non-sinusoidal: long elevations of $n_e$ (as opposed to area of first bunch) are alternated by short decreases of $n_e$. The bunches are located in the regions of the elevations of $n_e$.

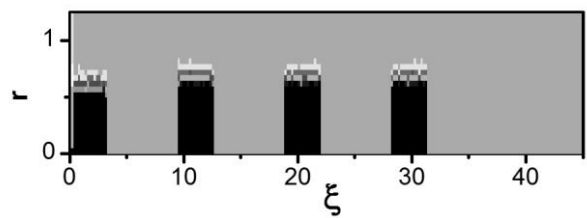

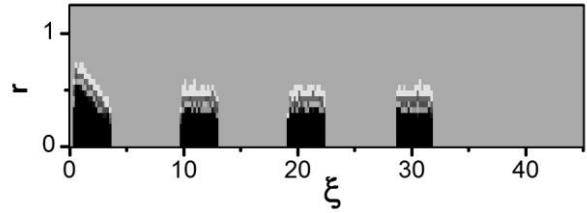

Fig. 3. The spatial (r, ξ) distribution of electron density $n_b$ of bunches before focusing (near the boundary of beam injection, z=3) and after focusing (into the plasma on the distance z=20 from the boundary of beam injection) for $\gamma_b=5$, $I_b=0.3\times10^{-3}$, $r_b=0.1$. Black corresponds to the maximum density, and gray– to zero

In linear approximation relation $E_r\sim\partial_r E_z$ for transversal and longitudinal wakefields is known. But in our case it is erroneous (One can see Fig. 1). It is obviously from Figs. 1-2 that positive and negative perturbations of $E_z$ are alternated, compensating each other. The latter leads to observed fact that along the final areas of dimension $\lambda/2$ of positions of bunches the longitudinal wakefield $E_z$ is compensated. But in these areas of positions of bunches the transversal force $F_r$ is finite $F_r\neq 0$.

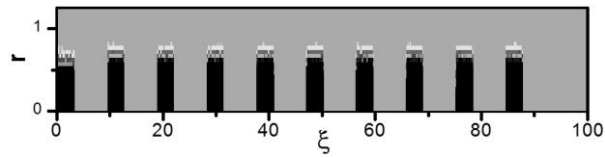

Fig. 4. The spatial distribution of electron density $n_b$ of ten bunches before focusing for z=3, $\gamma_b=1000$, $I_b=0.3\times10^{-3}$, $r_b=0.1$. Black corresponds to the maximum density, and gray – to zero

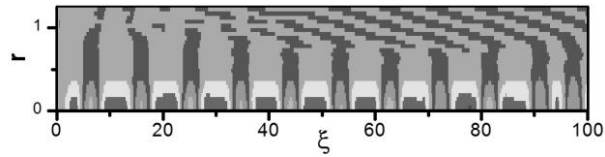

Fig. 5. Plasma electron density $n_e$ in wakefield, excited by ten bunches for z=3, $\gamma_b=1000$, $I_b=0.3\times10^{-3}$, $r_b=0.1$

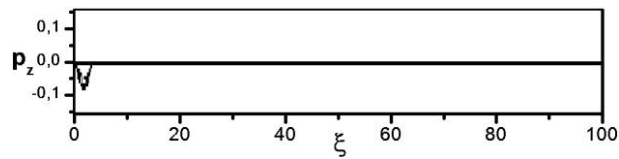

Fig. 6. Change of longitudinal momentum of bunches $P_z$ at wakefield excitation for z=80, $\gamma_b=1000$, $I_b=0.3\times10^{-3}$, $r_b=0.1$, $p_z=P_z-1$

Let us show that similar behavior and dependencies are observed for another number of bunches. In particular, we simulate the case of ten bunches (one can see Fig. 4). Similar to the case of four bunches, the first bunch current is in two times smaller than the current of the following bunches of sequence $I_1=I/2$, $I_i=I$, i=2, 3, 4 ..., spatial dimension from one bunch to another coincides with $1.5\lambda$. One can see the plasma electron density $n_e$ perturbation, excited longitudinal wakefield $E_z$, radial wake force $F_r$, magnetic field $H_\theta$, average wakefield $\langle E_z\rangle$, momentum of bunches $P_z$ in Figs. 4-8. One can see that behavior and dependencies are identical to the case of four bunches. Namely, Fig. 7 demonstrates that in the case of ten bunches also $\langle E_z\rangle$ for only bunch on front of sequence is finite and hence only bunch on front of sequence is decelerated (Fig. 6).

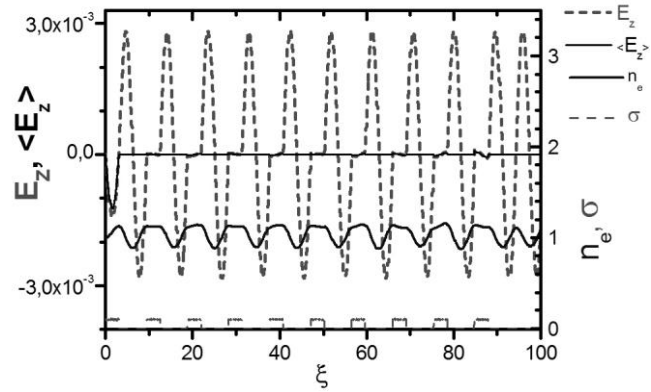

Fig. 7. On-axis plasma electron density $n_e$ (------), emittance of bunches $\sigma$ (········), on-axis longitudinal wakefield $E_z$ (- - -) and $\langle E_z \rangle$ (------) coupling rate of bunch with wakefield $E_z$ for z=3, $\gamma_b$=1000, $I_b$=0.3×10$^{-3}$, $r_b$=0.1

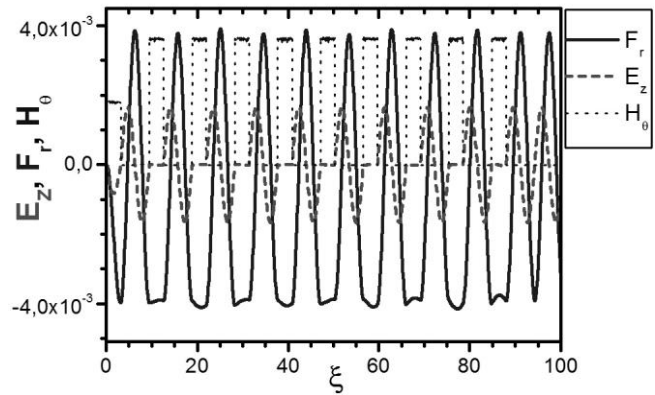

Fig. 8. Off-axis longitudinal wakefield $E_z$ (- - -), off-axis radial wake force $F_r$ (----) and off-axis magnetic field $H_\theta$ (······) for z=3, $\gamma_b$=1000, $I_b$=0.3×10$^{-3}$, $r_b$=0.1. $E_z$, $F_r$, $H_\theta$ have been calculated for radius r=$r_b$

Again one can see (Fig. 8) that only 1-st bunch is in finite $E_z\neq 0$. Other bunches are in zero longitudinal electrical wakefield $E_z$=0. Radial wake force $F_r$ in regions, occupied by bunches, is finite (Fig. 8).

## CONCLUSION

It has been shown that in considered conditions all relativistic positron bunches unlike a bunch at the front of sequence are focused equally and homogeneously (Fig. 3) similar to electron bunches. For this it is necessary that wavelength coinciding with double longitudinal dimension of bunches, the first bunch current is in two times smaller than the current of the following bunches of sequence, spatial dimension from one bunch to another coincides with 1.5λ. From results of simulations one can conclude that only bunch on the front of sequence interacts with wakefield. All next bunches are in areas, where longitudinal wakefield equals zero. Therefor only bunch on the front of sequence excites wakefield. The next bunches do not excite wakefield. Hence value of wakefield does not increase along sequence. Transversal force in areas of bunches is the same and homogeneous along bunches.